\documentclass[lettersize,journal]{IEEEtran}
\usepackage{amsmath,amsfonts}
\usepackage{algorithmic}
\usepackage{algorithm}
\usepackage{array}
\usepackage[caption=false,font=normalsize,labelfont=sf,textfont=sf]{subfig}
\usepackage{textcomp}
\usepackage{stfloats}
\usepackage{url}
\usepackage{booktabs}
\usepackage{colortbl}
\usepackage{tabularray}
\usepackage{verbatim}
\usepackage{graphicx}
\usepackage{cite}
\usepackage{color}
\usepackage{colortbl}
\usepackage[justification=centering]{caption}
\usepackage{ragged2e}
\definecolor{Gray}{gray}{0.9}
\definecolor{LightCyan}{rgb}{0.88,1,1}
\usepackage[dvipsnames]{xcolor}
\definecolor{blue}{HTML}{95bddc}
\definecolor{lightblue}{HTML}{c2d1e5}
\definecolor{orange}{HTML}{fe793d}
\definecolor{red}{HTML}{fb4c1f}
\definecolor{lightbrown}{HTML}{b71a3b}
\definecolor{brown}{HTML}{7e0f12}
\definecolor{bluegrey}{HTML}{a1afc9}
\definecolor{red1}{HTML}{E85642}
\definecolor{red2}{HTML}{C00000}
\definecolor{grey}{HTML}{767C7B}
\definecolor{bluetype1}{HTML}{1280B0}
\definecolor{bluetype2}{HTML}{17A1DE}
\definecolor{bluetype3}{HTML}{25537D}
\definecolor{yellow}{HTML}{f2f2b0}
\definecolor{color1}{HTML}{a0d8ef}
\definecolor{color2}{HTML}{c1e4e9}
\definecolor{color3}{HTML}{eaf4fc}

\begin{document}

\title{Self-Evolving Wireless Communications: A Novel Intelligence Trend for 6G and Beyond}

\author{Liangxin~Qian,~\IEEEmembership{Graduate Student Member, IEEE},
        Ping~Yang,~\IEEEmembership{Senior Member, IEEE},
        Jun~Zhao,~\IEEEmembership{Member, IEEE},
        Ze~Chen,
        Wanbin Tang
\thanks{L. Qian and J. Zhao are with the School of Computer Science and Engineering at Nanyang Technological University (NTU), Singapore. (e-mail: qian0080@e.ntu.edu.sg, junzhao@ntu.edu.sg). P. Yang, Z. Chen, and W. Tang are with the National Key Laboratory of Science and Technology on Communications, University of Electronic Science and Technology of China, 611731, Sichuan, China. (e-mail: yang.ping@uestc.edu.cn, 202222220229@std.uestc.edu.cn, wbtang@uestc.edu.cn).}
}

\markboth{}%
{}

\maketitle

\begin{abstract}
Wireless communication is rapidly evolving, and future wireless communications (6G and beyond) will be more heterogeneous, multi-layered, and complex, which poses challenges to traditional communications. Adaptive technologies in traditional communication systems respond to environmental changes by modifying system parameters and structures on their own and are not flexible and agile enough to satisfy requirements in future communications. To tackle these challenges, we propose a novel self-evolving communication framework, which consists of three layers: data layer, information layer, and knowledge layer. The first two layers allow communication systems to sense environments, fuse data, and generate a knowledge base for the knowledge layer. When dealing with a variety of application scenarios and environments, the generated knowledge is subsequently fed back to the first two layers for communication in practical application scenarios to obtain self-evolving ability and enhance the robustness of the system. In this paper, we first highlight the limitations of current adaptive communication systems and the need for intelligence, automation, and self-evolution in future wireless communications. We overview the development of self-evolving technologies and conceive the concept of self-evolving communications with its hypothetical architecture. To demonstrate the power of self-evolving modules, we compare the performances of a communication system with and without evolution. We then provide some potential techniques that enable self-evolving communications and challenges in implementing them.
\end{abstract}


\section{Introduction}
In recent years, wireless communication has been developing by leaps and bounds. While the fifth generation (5G) wireless services have been valuable, it is time to look forward to the sixth generation (6G) blueprint. In 6G, 10 million devices per square kilometer will be supported with seamless connections, and individual data rates will reach 100 Gbps. Moreover, 6G is expected to have much broader application scenarios, such as autonomous driving, virtual reality (VR)/augmented reality (AR), and smart cities. These emerging applications will lead to massive amounts of data that make manual processing very difficult or even impossible. As a result, automation must take center stage in the future~\cite{saad2019vision}.

\subsection{Limitations of Adaptive Communications}
In current wireless communications, adaptive technology is one of the main solutions to minimize human involvement and enhance automation. It can adjust system parameters to improve communication quality according to different environments, such as channels and interference. It has been widely used in the fourth generation (4G) for resource allocation and network optimization\cite{ku2014resource}. The optional modulation coding scheme (MCS) is added as a set of user-selectable waveform parameters in 5G. It is still very difficult to adapt links to meet diverse constraints and requirements, including rate, delay, and reliability. In 6G, more diverse types of terminals will work in intelligence-to-intelligence communication scenarios, requiring timely information on the terminal, base station, and network sides. Furthermore, the 6G network will be changed from signal-data content awareness into scenario-semantic awareness to support the connected intelligent devices. Traditional communication networks simply address the services around people, such as voice, image, and video. Machine-to-machine communications are expected to be dominant in 6G. Therefore, more sensing-fusion-decision-data-knowledge-based learning is needed to realize various tasks, such as autonomous collaboration of different types of robots/unmanned systems from different vendors. At this point, the purpose of the future network is to connect multiple intelligence together rather than simply communications. In adaptive communications, intelligence is mainly centralized or semi-centralized, which can't adapt to the future fully distributed terminal communication scenarios \cite{darwish2021vision}. Therefore, network architecture, resource management, and physical layer transmit technologies in future communication will be inherently intelligent and self-evolving, which is different from previous adaptive systems. In Table \ref{tab:comparison}, we present the comparisons between adaptive communications and future communications.
\begin{table*}\normalsize
    \centering
    \begin{tabular}{|l|p{6.5cm}|p{6.5cm}|}
        \hline
        \rowcolor{color1} \textbf{Comparative aspect} & \textbf{Adaptive communications} & \textbf{Future communications} \\
        \hline
        \cellcolor{color2} Perception & \cellcolor{color3} Signal-data content awareness & \cellcolor{color3} Scenario-semantic awareness \\
        \hline
        \cellcolor{color2} Intelligence deployment & \cellcolor{color3} Centralized or semi-centralized & \cellcolor{color3} Fully distributed \\
        \hline
        \cellcolor{color2} Adaptation & \cellcolor{color3} Choose policies from pre-designed policy databases & \cellcolor{color3} Not limited to fixed policy databases. New policies can be generated by genetic algorithms or other evolutionary algorithms \\
        \hline
        \cellcolor{color2} Communication & \cellcolor{color3} Human-machine system with the involvement of human & \cellcolor{color3} Machine-machine communication will be dominant and without the involvement of human \\
        \hline
        \cellcolor{color2} Machine learning & \cellcolor{color3} Lots of domain expertise and human intervention are needed & \cellcolor{color3} Sensing-fusion-decision-data-knowledge-based and no human intervention \\
        \hline
        \cellcolor{color2} Task diversity & \cellcolor{color3} Simple services, e.g., voice, image, and video & \cellcolor{color3} Various novel tasks, e.g., collaboration of robots/unmanned systems from different vendors, AI-generated content, and Metaverse.\\
        \hline
        \end{tabular}
\caption{Comparisons between adaptive communications and future communications}
\label{tab:comparison}
\end{table*}

\subsection{Development of Self-Evolving Technologies}
The term ``self-evolution'' first appeared in \cite{hely1997new} in 1997. Compared to the sparse distributed memory which is a static and inflexible system, the authors removed some problematic features of it and directly learned input patterns to achieve robustness and self-evolution. However, the mechanism and process of self-evolution were not clearly presented in \cite{hely1997new}. The meaning of ``self-evolution" was first elaborated in 1998 in \cite{dittrich1998self}. Self-evolution there refers to an evolutionary process in a system where the evolutionary behavior was carried out by the individuals of the system itself. The authors explored self-evolution in a binary string system in extinction, emergences of an organization, exploration-innovation, and evolution processes. Their work was motivated by chemical reaction dynamics, where molecular collisions and interactions produce new molecules that form metabolic networks. Instead of using mutation, selection, or recombination operations, a series of logical operations, e.g., AND, OR, NOT, EXOR, and EQ, and reactor algorithms were utilized in \cite{dittrich1998self}, where evolutionary phenomena were observed. A self-evolving knowledge base was proposed in 2005 \cite{barlas2005self}, and it indicates the ability to adjust the system configuration to improve its performance over time. It was originally developed for military systems that exploit a dynamic knowledge base for adaptation to address various tasks, such as self-learning, self-management, and self-diagnosis. The characteristics and structure of self-evolving systems were first shown in \cite{barlas2005self}. These work laid the basic prototype of the self-evolving technologies.

There are various evolutionary algorithms according to the Darwinian theory of evolution and the main branches of them are genetic algorithms, evolution strategies, evolutionary programming, generic programming, and classifier systems \cite{zhang2013swarm}. These genetic algorithms can be used to search for optimal or sub-optimal solutions in specific optimization problems. A self-evolutionary unicast routing protocol was proposed in \cite{tian2017microbial}, where the cellular attractor selection mechanism was used to select the next hop until the best routing path was found. Self-evolution is reflected in the adaptability and robustness of this protocol. Ye~\emph{et~al.}~\cite{ye2016agent} proposed a self-evolving service composition approach to deal with the evolving environments rapidly over time. It comprised five stages: service discovery, candidate selection, service negotiation, task execution, and network evolution. They were jointly optimized to make this service composition approach more efficient.
In \cite{chang2018deep}, a deep self-evolution clustering was proposed to jointly learn representations and cluster data. The proposed self-evolution mechanism is reflected in the pair-wise patterns selection. With the progress of learning, the algorithm gradually chose those patterns with maximum likelihood to improve the cluster accuracy and refine the model. An alternating iterative optimization algorithm, self-evolution clustering training, was also studied in \cite{chang2018deep} to select more accurate pair-wise patterns for training. Simulation results had shown superior performance compared with other cluster algorithms. Apart from these, an algorithm called deep deterministic policy gradient (DDPG) was utilized in \cite{chen2021dynamic}. The network of DDPG is trained off-policy with samples from the replay buffer to minimize sample correlation. At each time step, the self-evolution of the network is accomplished by updating the samples in the replay buffer when the buffer reaches its capacity.
In multi-task optimization problems, the uncertainty and challenges may increase, which can cause negative effects on optimization performance. A multisource knowledge transfer mechanism was proposed in \cite{liang2021evolutionary} to mitigate the effects. Based on the evolution experience, it estimated the likelihood of applying knowledge transfer adaptively and balanced the self-evolution in each task and knowledge transfer within multiple tasks. A novel self-evolving and transformative (SET) protocol was proposed in \cite{cai2022self} to support various services in 6G. The decomposition of feasible solutions can provide a library of control blocks to solve a wide range of problems. When an existing function cannot solve a new situation, new control functions can be created. The expanding function library causes SET protocols to self-evolve. The work presented in \cite{darwish2021vision} introduces a novel self-evolving network framework based on the intelligent vertical heterogeneous network architecture. The proposed framework addresses conflicts and manages coordination among multiple automated network entities while satisfying Quality of Experience (QoE) requirements of diverse user equipment (UEs). By leveraging artificial intelligence and machine learning, the framework enables dynamic and intelligent management of heterogeneous networks in an integrated network. 


\begin{figure*}[htbp]
    \centering
    \includegraphics[scale=0.5]{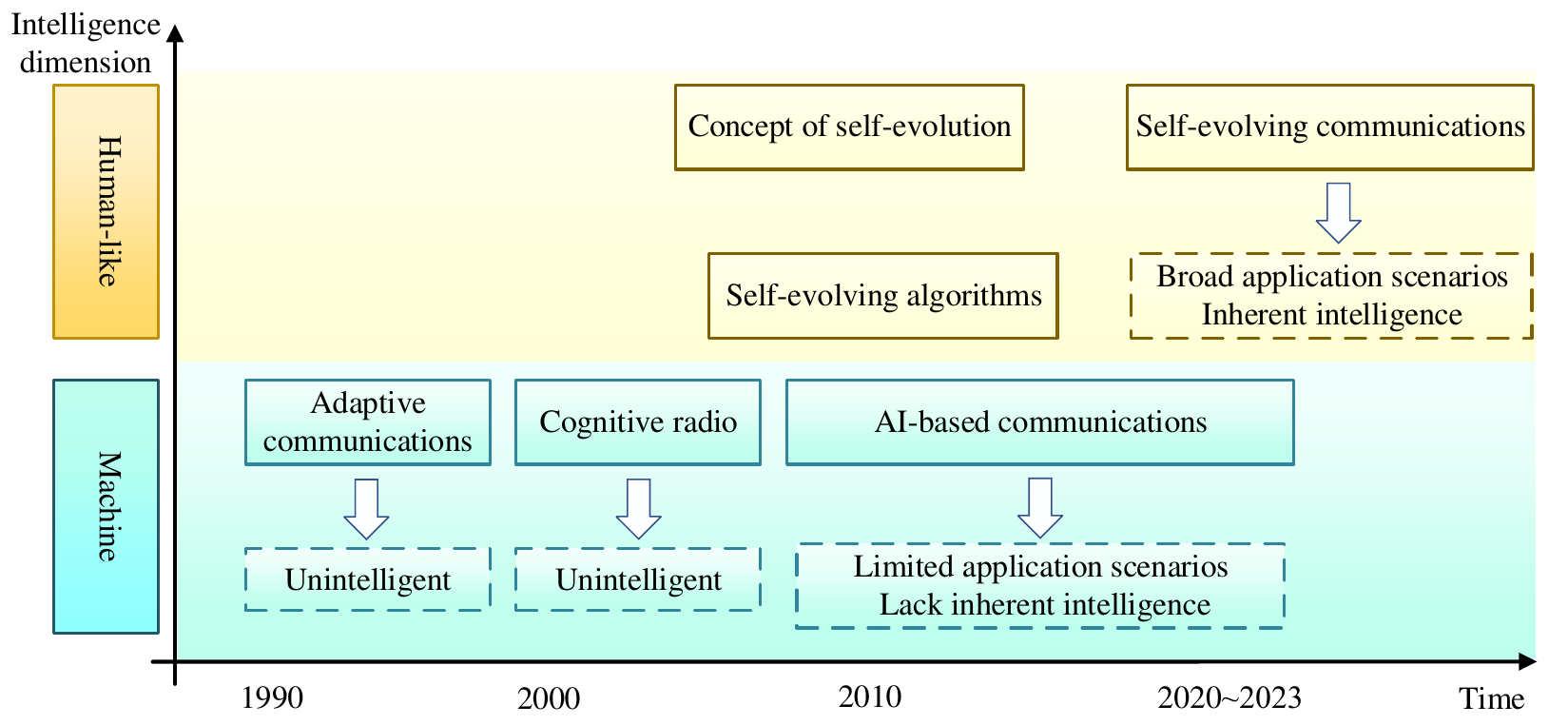}
    \caption{Self-evolving communication systems versus other communication systems.}
    \label{fig2}
\end{figure*}
To sum up, the above self-evolution-related work has a common mechanism: first, detect the new environment or service, then select the individuals who achieve the best performance in the new environment or service, and then replicate them, so as to achieve system self-evolution. Next, we will extend the concept of evolutionary communications based on this idea.

\subsection{Concept of Self-Evolving Communications}
To develop self-evolving communication systems, let us start with the basic meaning of evolution. Originally, evolution refers to the gradual change, development, and transition of things from one state to another, usually towards the direction of being more advanced, more complex, and more perfect. For example, we can regard the source code, channel code, modulation, and demodulation as the ``genes'' of a communication system. The process of learning and updating communication knowledge from the environment can thus be regarded as a process of ``genetic mutation'', ``genetic variation'', and evolution. Gradually, the genome of a self-evolving communication system will become entirely new and can better adapt to new environments.

For self-evolving communications, knowledge acquired from the environment plays an important role in the evolutionary process. To better illustrate the process of knowledge acquisition, we borrow some concepts from Szpankowski \cite{szpankowski2018frontiers}. The characteristic parameters in communication systems, such as the distribution and model of the signal or noise, are extracted from the sampled signal data and synthesized into valuable information. The learning algorithms then aim to learn some actionable insights from this information and integrate these insights into knowledge.

Based on the above discussion, \emph{a self-evolving communication system can adapt to different unknown environments, conduct cognitive reasoning, and continuously update the knowledge base to improve communication performance over time.} It is typically unsupervised.

\subsection{Different from Existing Systems}
In Fig. \ref{fig2}, we compare self-evolving communication systems with other communication systems. In addition to the adaptive technologies previously discussed, cognitive radio is also capable of sensing the environment and adapting its transmission parameters to smartly allocate resources and improve communication performance \cite{gavrilovska2013learning}. Cognitive radio only has some degree of automation and adaptation. It still requires manual assistance and lacks full intelligence.

In the past 20 years, machine learning, especially deep neural networks, has flourished and has led to many AI-based cognitive radio or adaptive intelligent communication frameworks. However, intelligent communication frameworks are still at the level of communication signal processing and network management and are unable to process massive data and acquire the information and knowledge to evolve communication systems and networks. Therefore, they cannot adapt to broad intelligent application scenarios in the 6G and beyond. More than that, current AI-based communication frameworks focus on observation and adaptation and lack endogenous learning and cognitive reasoning capabilities \cite{gavrilovska2013learning}. For example, AI-based adaptive technologies can only use limited information to choose fixed policies in terms of decision-making. However, the self-evolving communication system proposed in this paper is able to cognitively recognize information about various types of intelligent application scenarios and can continuously learn, reason, accumulate, and update environment and policy knowledge. On top of that, it can modify and improve the policy base for different environments through contention.

\section{Self-Evolving Wireless Communications}
In this section, we present a hypothetical model of a self-evolving communication system comprising data, information, and knowledge layers. We compare the performance of communication systems with and without self-evolving modules. Using the example of the Extreme Learning Machine (ELM) technique, we introduce the self-adaptive evolutionary ELM (SaE-ELM) that improves feature extraction and network parameter optimization. Simulations demonstrate the superior performance of SaE-ELM in terms of bit error rate (BER). Additionally, we propose a self-evolving and Q-learning-based (SE-QL) algorithm for joint space-frequency rendezvous in UAV systems, which outperforms traditional Q-learning methods in terms of efficiency and convergence. These results highlight the potential of self-evolving communication systems to adapt, evaluate strategies, update knowledge, and achieve automation in future wireless communication networks.
\subsection{Hypothetical Model}
\begin{figure*}[htbp]
    \centering
    \includegraphics[scale=0.5]{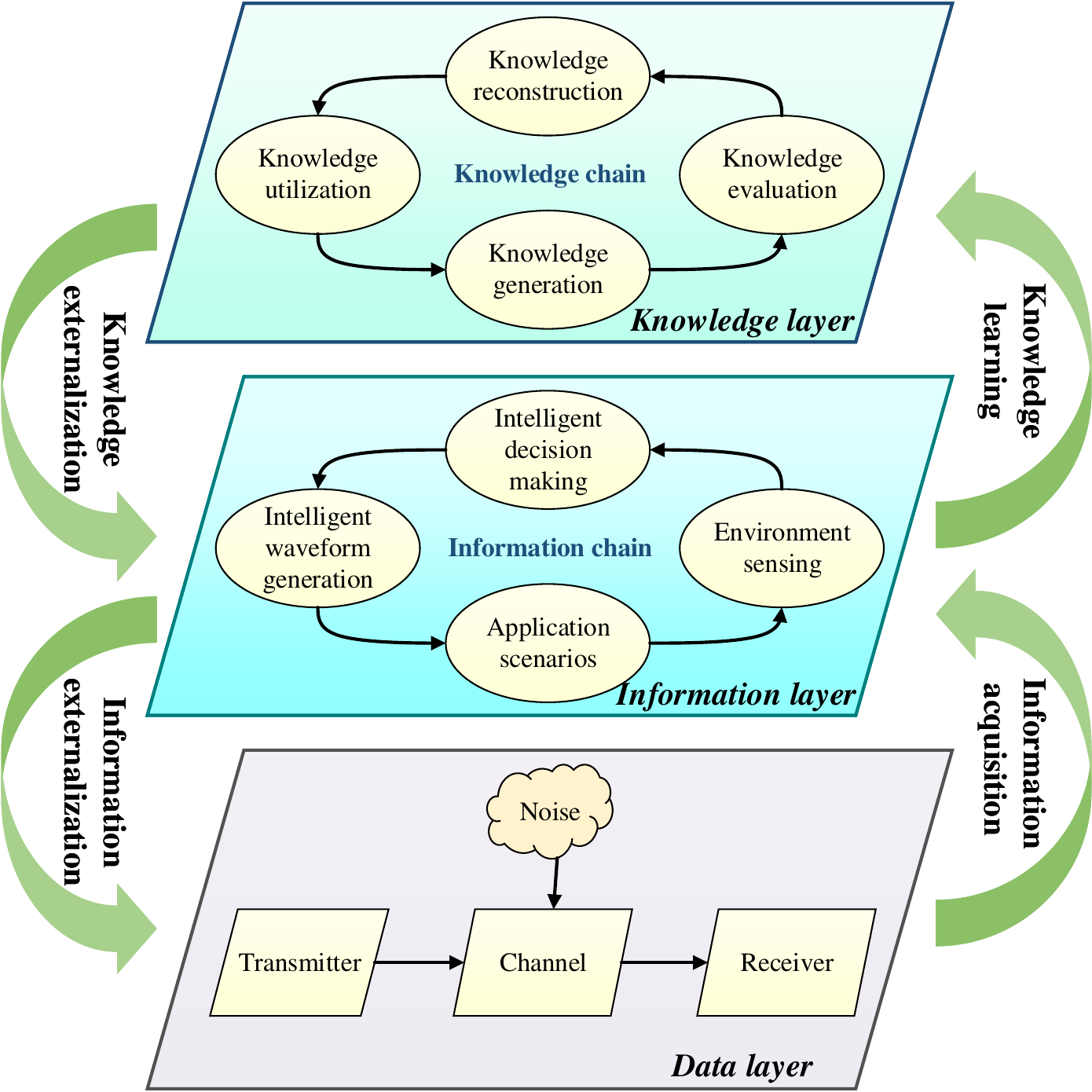}
    \caption{A conceptual diagram of a self-evolving communication system.}
    \label{fig1}
\end{figure*}

In Fig. \ref{fig1}, we illustrate a self-evolving communication system. It includes the data layer, the information layer, and the knowledge layer. The data layer consists of conventional communication modules, such as transceiver, channel, and noise. The information layer includes four modules: application scenarios, environment sensing, intelligent decision-making, and intelligent waveform generation. The knowledge layer consists of four modules: knowledge generation, knowledge evaluation, knowledge reconstruction, and knowledge utilization. Information acquisition, knowledge learning, knowledge externalisation, and information externalization are accomplished through interaction among the three layers.

As shown in Fig. \ref{fig1}, the data layer includes conventional communication modules and addresses the problems of reliable transmission of signals by using information theory and signal theory. The data layer also allows us to acquire information about environments, signal waveforms, etc., from various application scenarios. The measured status data from the communication modules will be fed into environment sensing modules, intelligent decision-making modules, and intelligent waveform generation modules in the information layer, where they are analyzed to obtain useful information. The circular process of acquiring information between these three modules and application scenarios is called the information chain. The obtained information is used in the learning phase of these modules.

The information layer modules cycle through learning iteratively and generate new knowledge for the knowledge layer. The knowledge layer conducts cognitive reasoning by evaluating the validity of the knowledge, adding new useful knowledge, or removing aged one through knowledge evaluation and reconstruction. The new knowledge at the knowledge layer will be utilized in the information layer and eventually the data layer for optimizing communication systems. As shown in Fig. \ref{fig1}, these processes are called knowledge and information externalization, respectively. The circular process from knowledge generation to knowledge utilization in the knowledge layer is called the knowledge chain. The feedback from the knowledge layer or knowledge externalization determines the new direction in which the communication system will evolve.

Based on the above discussion, a self-evolving communication system may briefly entail the following features:\\
\emph{$\bullet$ Self-learning}: Through cognitive reasoning, acquiring critical knowledge that affects system performance and can adjust system parameters timely for improving performance under different and probably new scenarios.\\
\emph{$\bullet$ Self-evaluation}: Evaluating the performance of the corresponding communication strategies and then determining a communication strategy that best suits the new environments.\\
\emph{$\bullet$ Self-management}: Dynamically updating knowledge in the knowledge base. Removing the knowledge that will lead to poor system performance or has a low probability of being used. Adding the new knowledge resulting from interaction with new environments for evaluation and application.\\
\emph{$\bullet$ Automatic}: The self-evolving communication system may initially be able to organize the data itself through some approaches. Or, for the sake of efficiency, to manually observe and process some data first and provide some useful knowledge to the system. But eventually, there should be little manual intervention in the system.

\subsection{Comparison of Communication Performance with or without Self-Evolving Modules}
To demonstrate the power of self-evolving modules, we compare the performances of a communication system with and without evolution. We first take the ELM technique as an example, which has a single hidden layer neural network (NN) that discovers relationships between input and output data by imitating human neurons. It not only has a relatively simple network structure but also needs no updating on network parameters iteratively. It works in two steps: feature extraction and mapping. Feature on the input data is extracted and the parameters between the input layer and the single hidden layer are assigned randomly. Then, the parameters between the single hidden layer and the output layer are calculated through matrix operations and finally are mapped to the output.

Due to the random parameter assignment in the first step, the effect of feature extraction cannot be well guaranteed. Therefore, the self-adaptive evolutionary extreme learning machine (SaE-ELM) has been proposed in \cite{cao2012self} to extract features more effectively by utilizing the self-adaptive differential evolution algorithm. The idea is to predefine a parameter set and a policy set. In each iteration, a policy and a parameter will be selected from the policy set and the parameter set to update the current parameters, finally making the network evolve continuously and obtain better performance.

\begin{figure}[!t]
    \centering
    \includegraphics[scale=0.52]{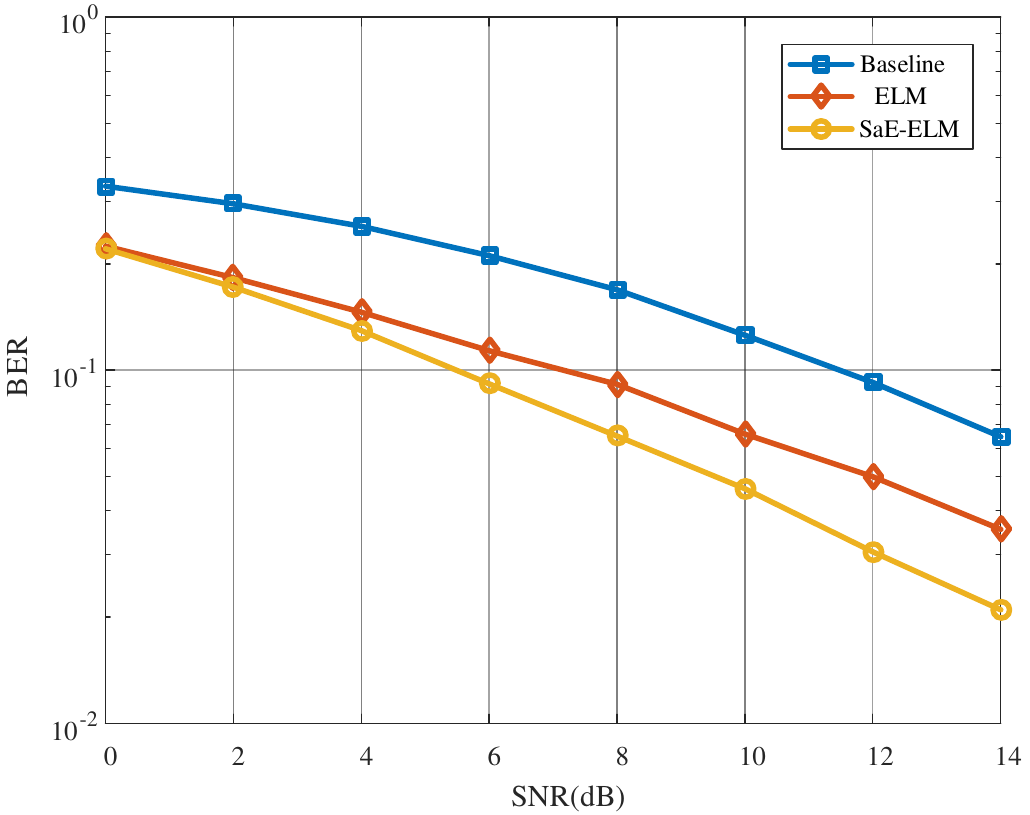}
    \caption{BER comparison of conventional ELM and SaE-ELM methods in Rayleigh channels.}
    \label{fig3}
\end{figure}

\begin{figure}[!t]
    \centering
    \includegraphics[scale=0.65]{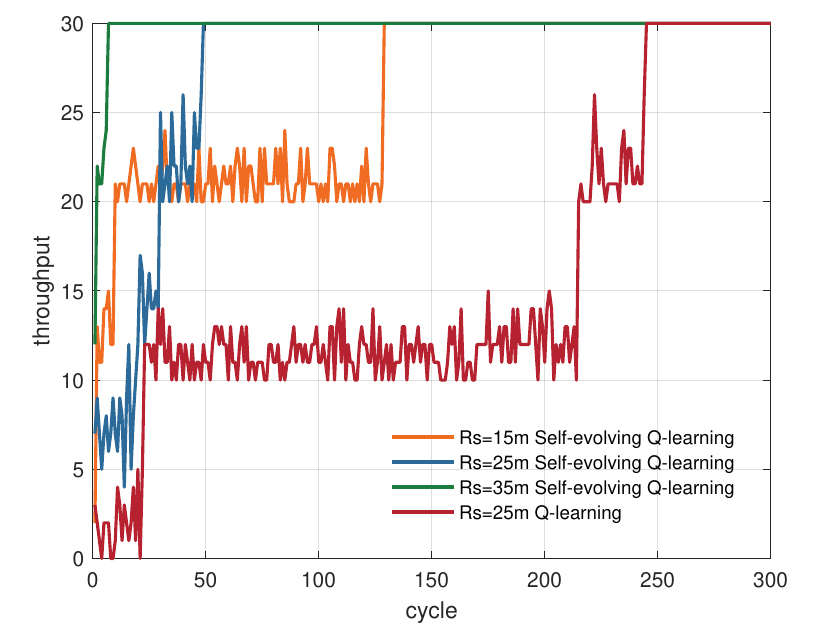}
    \caption{Throughput comparison of Q-learning and SE-QL methods in UAV rendezvous.}
    \label{fig4}
\end{figure}

Consider a point-to-point communication system with one single antenna at both transmitter and receiver and with binary phase shift keying (BPSK). The channel is with Rayleigh fading. The SaE-ELM and ELM are adopted as detection modules, which represent communication systems with and without self-evolving elements, respectively. At each signal-noise ratio point, we generate 100,000 training data sets, including channel matrix as input and BPSK symbols as output, of which 70\% are used for training and 30\% for testing. We adopt four parameter-changing strategies in \cite{cao2012self} and set $\xi$ to 0.01, which is a small positive constant to avoid the possible null success rate. The number of hidden neurons of both SaE-ELM and ELM is 1000, and the batch size is 100. The number of iterations of SaE-ELM is set to 50, and the size of the parameter set is set to 100. We assume a classical detection method (Zero-forcing) as a baseline without machine learning. From Fig. \ref{fig3}, there is nearly a 6 dB bit error rate (BER) performance gap between the baseline and SaE-ELM method. The BER of the SaE-ELM method with the adaptive differential evolutionary algorithm is about 2 dB higher than that of the ELM method. The above results indicate that the SaE-ELM method can effectively avoid the limitations of parameter control in the ELM method and further improve the generalization performance of the network by introducing the adaptive differential evolutionary algorithm to optimize the network parameters.

In addition to communication system detection performance, we also investigate the rendezvous performance of unmanned aerial vehicle (UAV) systems with and without self-evolution in this work. To achieve joint space-frequency rendezvous for a swarm of UAVs, a Q-learning-based algorithm that directly optimizes the policy actions of the UAVs using reinforcement learning techniques is proposed. The algorithm maintains and updates a value table or value function through the Bellman equation, whereby the action with the greatest value is selected.

However, due to the difficulty of Q-learning in handling massive or continuous state and action spaces, the algorithm may fail to converge. To address this limitation, we propose a self-evolving and Q-learning-based (SE-QL) joint space-frequency rendezvous method for UAV swarms. The method uses the state at the end of each Q-learning round as part of the initial population, randomly generates states as other individuals, and then performs selecting, copying, exchanging, mutation, and updating. The individual with the best performance in the population is selected using a genetic algorithm and used as the initial state for the next round of Q-learning, leading to improved space-frequency rendezvous performance and accelerated convergence compared to the classic Q-learning method without evolution.

The proposed method is evaluated on a simple UAV rendezvous scenario where multiple UAVs form clusters in an open two-dimensional region, and UAVs within the same cluster need to rendezvous at the same frequency to establish communication. We randomly generate the initial positions of 15 UAVs and set the number of channels to 3, the learning rate $l_r$ of Q-learning to 0.85, and the decay factor to 0.99. Further, we use the Q-learning method without self-evolution as the baseline and compare it with the Q-learning method with self-evolution. In Fig. \ref{fig4}, results indicate that for a sensing range of 25m, the Q-learning method with self-evolution uses 80\% fewer cycles than the baseline to reach convergence, resulting in improved algorithm efficiency. Additionally, for a sensing range of 35m, the algorithm converges more rapidly. Notably, for a sensing range of 15m, the convergence rate of the baseline is still twice as fast, highlighting the robustness of the proposed algorithm. Our findings demonstrate that the SE-QL method is effective in circumventing the issue of poor algorithm convergence in larger state spaces and can better leverage the airspace motion capability to manipulate the network topology and improve the efficiency of UAV swarm frequency rendezvous.

\section{Promising technologies}
\begin{figure*}[htbp]
    \centering
    \includegraphics[scale=0.37]{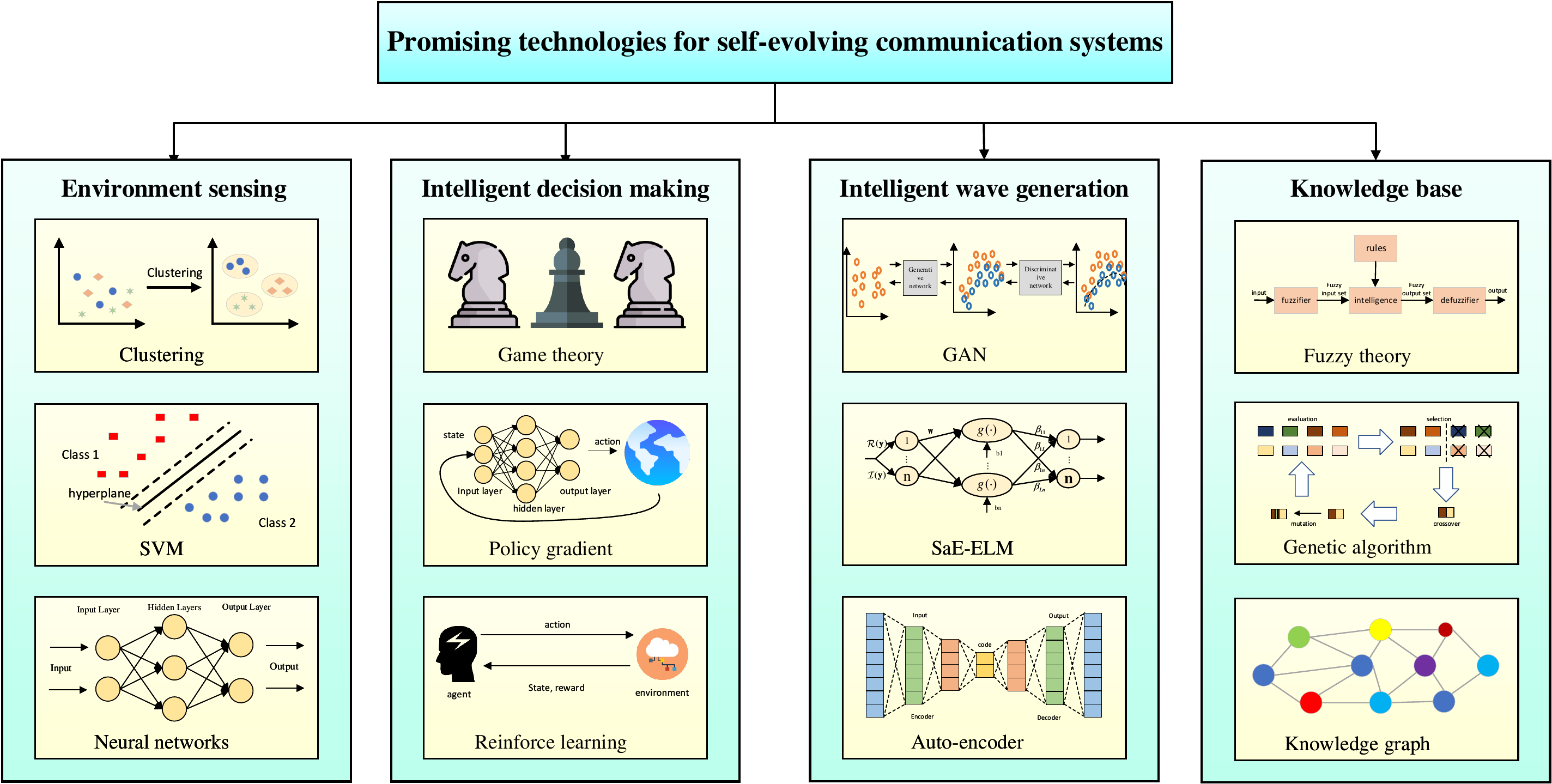}
    \captionsetup{justification=RaggedRight}
    \caption{Promising technologies corresponding to each module. In the information layer, classification algorithms in environment sensing are used to extract information in different environments or semantic scenarios, algorithms in intelligent waveform generation are used to handle optimization problems of receiving transmitters and waveforms, and algorithms in intelligent decision-making are used to assist in various optimization problems and communication resource allocation problems, etc. The algorithms for the knowledge base are used for the construction and management of the knowledge base.}
    \label{fig5}
\end{figure*}

Over a couple of decades, there have been many self-evolving related techniques in the context of cognitive radio and autonomous communication, as shown in Fig. \ref{fig5}. For example, various types of machine learning problems and algorithms in cognitive radio are summarized in \cite{bkassiny2012survey}; deep learning techniques have been applied in the physical layer in \cite{o2017introduction}, trying to regard communication as an end-to-end reconfiguration optimization task; domain knowledge has been integrated into deep learning models in \cite{she2021tutorial}. The above works and the references therein can offer readers a trail to follow and a deeper understanding of these techniques. In this section, we will discuss some potential technologies that can be applied to self-evolving communication systems.

\textbf{Environment Sensing}: Proper knowledge of the environment is a prerequisite for the operation of a self-evolving communication system. Traditional environment sensing techniques involve extracting features and classifying environmental signals for cognition. Currently, there has been a lot of research using machine learning methods to solve environment sensing problems. For example, we can use support vector machines (SVM) (a generalized linear classifier that performs binary classification of data) or NN to classify different communication scenarios or signals. These methods require manual training of machine learning models in advance. The classification performance depends heavily on the available data set and signal features. Despite this, both NN and SVM need to be trained in different environments. Dirichlet process mixture model (DPMM) \cite{bkassiny2012survey}, a clustering algorithm that divides a collection of physical or abstract objects into multiple classes composed of similar objects, is fully autonomous and does not require any prior knowledge of known signals or the number of different signal classes. Furthermore, the non-parametric nature allows the DPMM to address an arbitrary number of clusters that can learn and act autonomously in any radio frequency environment.

\textbf{Intelligent Decision-Making}: Intelligent decision-making is a key part of a self-evolving communication system for executing self-evolving processes. The correct self-evolving behaviour results from the correct decision-making.

According to whether the state of the environment can be fully observed, we can classify the environment into Markov decision processes and partially observable Markov decision processes. For Markov decision processes, we can repeatedly interact with the environment by means of reinforcement learning (RL), receive feedback or rewards from the environment after executing a model/policy, obtain state information, and update the model's parameters. This allows the system to adjust decision policy according to the environment. For multiple-agency RL, different agencies interact directly with the environment in a distributed way, helping intelligent agencies to find the best possible decisions \cite{bkassiny2012survey}.

For partially observable Markov decision processes, the policy gradient method is more efficient than RL in that it can find optimal strategies directly in the strategy space. The policy vector can be continuously updated to find the overall or local optimal policy. Beyond this, game theory is also an effective mathematical tool to solve the problem of multiple intelligent agents competing for resources by exploiting the behaviour of competing individuals to optimize their strategies. It has been widely used in communication networks, particularly for resource allocation in competitive environments \cite{bkassiny2012survey}.

\textbf{Intelligent Waveform Generation}: Intelligent waveform generation modules apply knowledge learned to generate waveforms. With the development of deep learning techniques in the physical layer of communication, communication modules are gradually becoming intelligent and automated. In addition to the SaE-ELM technique mentioned before, the trainable point-to-point auto-encoder (AE) mentioned in \cite{o2017introduction} can perform encoding and decoding operations without any prior knowledge. Not only can communication modules be modeled by a neural network, but also can channel effects. In fast-changing environments where accurate channel information is impossible to be obtained, generative adversarial networks (GAN) are proposed to efficiently represent channel effects. They consist of two networks to generate data that resembles real channel data and to distinguish whether a data set is real or not \cite{simeone2018very}. In addition, RL can be utilized to enhance interaction with the environment and assist their optimization.

\textbf{Knowledge Bases}: The knowledge base is like the brain of a self-evolving communication system. Genetic algorithms can search for optimal solutions by simulating natural evolutionary processes. Traditionally, they are often used to design fuzzy logic variables and fuzzy decision rules based on the empirical data of the system. Genetic algorithms can be used to establish the knowledge base from the obtained information \cite{fazzolari2012review}. Information data may affect the system's performance positively or negatively. A better understanding of the relationship between these concepts will help communication systems evolve in a better and healthy direction. Knowledge graphs and relational machine learning are effective tools to predict these facts \cite{nickel2015review}.

\section{Challenges and perspectives}
\begin{figure*}[htbp]
    \centering
    \includegraphics[scale=0.5]{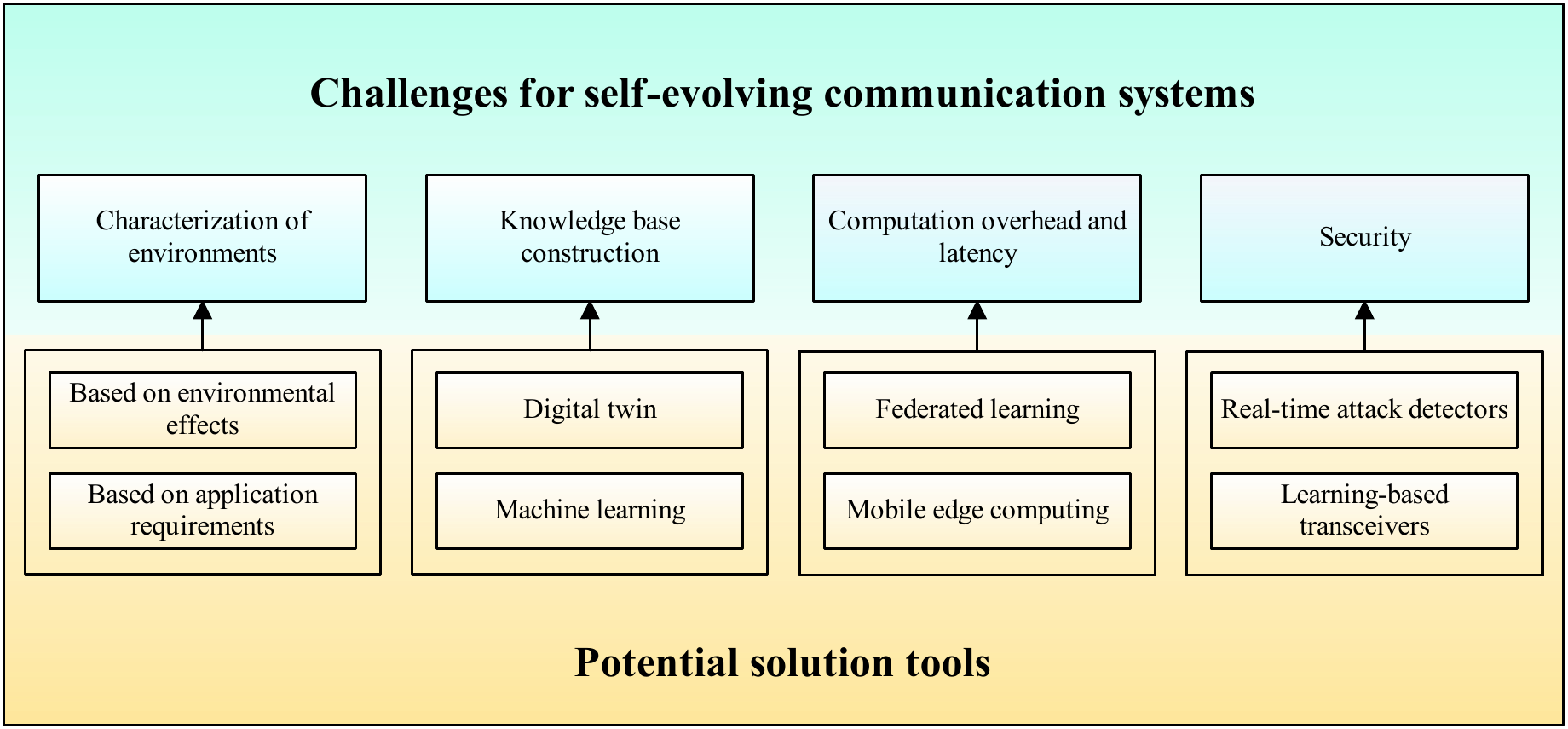}
    \caption{The challenges and corresponding potential solution tools for self-evolving communication systems.}
    \label{fig6}
\end{figure*}

Currently, the research on self-evolving communication systems is still in the preliminary stage. Although there have been some potential technologies for self-evolving communication systems, as we have discussed above, there are still many challenges ahead, as summarized in Fig. \ref{fig6}.
\subsection{Characterization of Environments} Characterization of environments refers to the representation of environments by some characteristic parameters. In the past, we usually characterize electromagnetic environments by the six domains, including space, time, frequency, energy, polarization, and modulation. However, they will not be enough for future communications. For example, the means for jamming will be more diversified and intelligent. Traditional six-domain characterization can hardly portray such a complex electromagnetic environment well in the future. In addition to the above, the future electromagnetic environment should also include signal density, variety, dynamism, and intense confrontation.

The classification and quantification method of the electromagnetic environment is in the standard IEC-TR-61000-2-5 proposed by the International Electrotechnical Commission (IEC). It begins with the electromagnetic environment by counting the values of specific measurement parameters (e.g., frequency band, signal rise time, voltage) based on different electromagnetic phenomena in different platforms. This solution focuses on environmental effects. From the point of view of application requirements, the cognition of electromagnetic signals can be divided into the signal level and information level macroscopically. At the signal level, parameters, such as time, frequency, motion, polarization, waveform, modulation, and baseband signal, can be analyzed to solve application requirements, such as spectrum management, signal identification and analysis, and feature measurement. At the information level, the information of the electromagnetic environment can be grasped by analyzing the encoding and encryption of signals so as to facilitate decision-making and solve the application requirements of information acquisition, and management control.

\subsection{Knowledge Base Construction} There are two approaches to establishing knowledge bases. First, a knowledge base can be designed by a group of experts or formed using known expertise \cite{nickel2015review}. Such a knowledge base is typically well-curated, professional, and capable of producing highly accurate results. However, it relies too much on experts or expert knowledge, and is hard to adapt to new environments, applications, and technologies. In real communication systems, we often encounter changing environments and a wide variety of data, including signal data, channel information, network routing information, etc., in which case expert knowledge may not be applicable. Knowledge bases can also be constructed by purely machine learning means (e.g., deep reinforcement learning (DRL) \cite{she2021tutorial}) based on real-time communication data, relevant policy evaluation, and real-time simulation results. Automatic knowledge base construction approaches cater to the future big data era and are gaining more and more attention. Since the DRL algorithm may suffer from the model mismatch, exploration risk, and slow convergence, digital twins can be utilized as a potential technique to mitigate these risks by creating a real-time high-fidelity digital visualization model of physical objects \cite{she2021tutorial}. In addition, with the help of relational machine learning, knowledge graphs can effectively discover new facts about physical objects or new relationships among objects, which can be utilized to construct the knowledge base \cite{nickel2015review}.

\subsection{Computation Overhead and Latency} As communication networks gradually evolve towards 6G, higher frequency spectrum (e.g., tera-hertz band) will be utilized, the data rate will reach tera-bit per second, the required end-to-end (E2E) delay will be less than one millisecond, system reliability requirements will be stricter. At the same time, a large number of intelligent transceiver devices will appear. As a result, the signal processing complexity will become higher, and optimization problems will be more elaborate. All the above issues demand the computing ability of future communication systems.

Mobile edge computing (MEC), a new computing platform, can provide services and cloud computing capabilities required by users on the wireless side. It is considered as a potential for ultra-reliable and low-latency communications (URLLCs) \cite{she2021tutorial}. MEC has lower delays than centralized cloud computing and may achieve one-millisecond latency with guaranteed reliability. Recently, distributed learning (DL) and federated learning (FL) have been investigated to mitigate the high information exchange overhead between the central controller and other base stations and the high computation overhead of each \cite{she2021tutorial}. In contrast to the above approaches, quantum computing has been also widely recognized as a key enabling technology for implementing complex computing systems and solving complex optimization problems because of its powerful large-scale parallel computation capability by using quantum information units \cite{saad2019vision}.

\subsection{Security} As the number of intelligent agents increases in the future, the methods for adversarial attackers become more intelligent and less easily detectable. Such new types of security intelligent problems (e.g., deep learning-assisted attackers) are called adversarial security. They act intelligently and simultaneously to evaluate the stability of the system, spreading false approximation information to make the system adopt the wrong policies. Current decision networks of intelligent entities at the physical and network layers appear vulnerable to such real-time decision perturbations. Also, adaptive modulation techniques are at the risk of indirectly exposing link information due to their need for channel state information or transmitter information. There are two possible solutions to this type of malicious adversarial attack: building a real-time attack detector and building a learning-based transceiver to resist malicious attacks and reduce their impact. For example, the automated learning-based transceiver, which does not require knowledge of the channel model, can undergo adversarial training \cite{carlini2017towards} to generate a more robust transmit network or decision network.

\subsection{Other Challenges} Even if the current supervised learning methods have achieved good results, they generally require a lot of labeling work. Moreover, they are less applicable to new categories (e.g., new communication devices) because they require the labels of the existing categories. In order to obtain faster adaptation in such cases, we can use few-shot or zero-shot learning, which aims to find historical tasks similar to the current one, and use meta-learning to obtain the initial values of the neural networks. It allows fast adaptation of the model with very few samples. In addition to this, we can use GANs to synthesize data artificially when very limited data is available in real systems or communications, such as image generation, signal generation, end-to-end latency, and channel gains \cite{she2021tutorial}.

\section{Conclusion}
Self-evolving communication systems represent a novel intelligence trend for 6G and beyond, enabling adaptive, intelligent, and autonomous networks that can continuously learn, reason, and improve communication performance. In this paper, we have explored the concept of self-evolving communications and its potential in addressing the limitations of current adaptive systems. By leveraging evolutionary algorithms, self-evolving communication systems can adapt to different unknown environments, conduct cognitive reasoning, and continuously improve their communication performances. The hypothetical model presented in this manuscript illustrates the integration of data, information, and knowledge layers in a self-evolving communication system, enabling self-learning, self-evaluation, self-management, and automation. To enable the deployment of self-evolving communications, we present potential technologies that can be applied to self-evolving communication systems. However, several challenges remain, including the need for efficient algorithms, handling massive data, addressing complex network scenarios, and ensuring robustness and security. Overcoming these challenges will require interdisciplinary research efforts and collaborations between academia, industry, and standardization bodies. As wireless communication continues to evolve, self-evolving communication systems hold great promise in enabling the vision of 6G and beyond, supporting diverse intelligent applications, and facilitating the seamless connectivity of multiple intelligence.
\bibliographystyle{IEEEtran}
\bibliography{reference}
\end{document}